# Video Game Development in a Rush:
# A Survey of the Global Game Jam Participants

Markus Borg
RISE Research Institutes of Sweden AB
markus.borg@ri.se

Vahid Garousi
Wageningen University, Netherlands
vahid.garousi@wur.nl

Anas Mahmoud
Louisiana State University, USA
mahmoud@csc.lsu.edu

Thomas Olsson
RISE Research Institutes of Sweden AB
thomas.olsson@ri.se

Oskar Stålberg
Plausible Concept, Sweden
info@oskarstalberg.com

**Abstract:**

Video game development is a complex endeavor, often involving complex software, large organizations, and aggressive release deadlines. Several studies have reported that periods of "crunch time" are prevalent in the video game industry, but there are few studies on the effects of time pressure. We conducted a survey with participants of the Global Game Jam (GGJ), a 48-hour hackathon. Based on 198 responses, the results suggest that: (1) iterative brainstorming is the most popular method for conceptualizing initial requirements; (2) continuous integration, minimum viable product, scope management, version control, and stand-up meetings are frequently applied development practices; (3) regular communication, internal playtesting, and dynamic and proactive planning are the most common quality assurance activities; and (4) familiarity with agile development has a weak correlation with perception of success in GGJ. We conclude that GGJ teams rely on ad hoc approaches to development and face-to-face communication, and recommend some complementary practices with limited overhead. Furthermore, as our findings are similar to recommendations for software startups, we posit that game jams and the startup scene share contextual similarities. Finally, we discuss the drawbacks of systemic "crunch time" and argue that game jam organizers are in a good position to problematize the phenomenon.

## 1 INTRODUCTION

Development of video games is a challenging task that involves the expertise of many skilled professionals from various disciplines including software engineering, art/media design, and business. Modern video games have enormous code-bases, and are comparable to large-scale conventional software systems, in terms of size and complexity. For example, the server-side code of a popular game named "World of Warcraft" [1] was reported to be about 5.5 MLOC [2]. The total revenue of the video game industry was about $30.4 billion in 2016 [3].

The pressure on game development to meet aggressive deadlines has contributed to a phenomenon known as "crunch time", i.e., long hours of overtime [4, 5]. Classic software engineering issues associated with game development includes requirements management, configuration management, and verification and validation [6]. Given the complexity of game software, game development companies are striving to develop game software with the highest quality and in the most effective and efficient manner. Thus, application of useful software engineering practices in this domain is important and could save costs and increase games' quality. Many studies have already studied application and usage of software engineering practices in game development, e.g., [6, 7].

A *game jam* is a hackathon for game development; i.e., participants gather in teams and develop games within a short time span (typically between 24-72 hours, and we do not consider longer game jams in this study) [8]. Despite the constant time pressure in game jam events, most teams do their best to develop and deliver (high quality) games. In this context, one might wonder how teams pursue a high-quality product, i.e., a successful game release.

There are plenty of established good game development practices, both from the game development scene and software engineering research [6]. But an important issue which has not been explored in the past is game development under extreme time pressure such as during a game jam. The general issue of software development time pressure has been studied in the past [9, 10], but it is important to study game development under time pressure, given the specific nature of game development and its difference to conventional software development [11], e.g., game development combines the work of teams covering multiple disciplines (art, music, acting, programming, etc.), and that engaging game play is sought after through the use of prototypes and iterations. More specifically, when developing a video game under time pressure, we raise and answer the following research questions (RQs) in this study:

- RQ 1: How do developers manage (capture and align) expectations ("requirements") of games that they develop, under time pressure?
  - RQ 1.1: How is initial gathering of game expectations conducted?
  - RQ 1.2: How is evolution and change of game expectations handled?
- RQ 2: How do game developers ensure that the game satisfies the expectations, under time pressure?
  - RQ 2.1: What type of Quality Assurance (QA) is done during game development?
  - RQ 2.2: What type of QA is done on the final product (game)?
- RQ 3: How do game developers apply established software and game development practices, under time pressure?
- RQ 4: Correlations of data: Are there any significant correlations among various factors in the survey, e.g., to

what extent are familiarity with lean development and perception of success correlated?

To answer the above RQs, we conducted a questionnaire-based survey with participants of the world's largest game-jam event, Global Game Jam (GGJ) in 2017. Our results in this paper is based on 198 responses that we received during that event, or the days after.

The remainder of this paper is structured as follows. Background and a review of the related work are presented in Section 2. We describe the research method and design of the opinion survey and its execution in Section 3. In Section 4, we present and analyze the results of the survey. In Section 0, we discuss the implications of the results, i.e., recommendations for future game jams, and possible generalization of the results to similar contexts. Finally, in Section 6 we conclude the paper and discuss future work directions.

## 2 BACKGROUND AND RELATED WORK

Video game development is different than development of conventional software, such as stand-alone applications and embedded systems. This is due to the specific nature of games, for example the major influence of graphics design and the elusive quality of fun. Software engineering of games is an active research area with many publications. A Systematic Literature Review (SLR) on software development processes for games [6], published in 2014, reviewed a total of 404 primary studies. Another 2010 SLR [7] on software engineering research for games reviewed 84 papers. We discuss a handful of papers from the large body of knowledge in this area.

One of papers in this area focused on what game developers test in their products [12]. The authors interviewed seven game development teams from different companies and studied how they test their products using the grounded-theory approach. Their results showed that game developers tend to focus on "soft" values such as game content or user experience, instead of more traditional objectives such as reliability or efficiency. The authors concluded that game developers have similar, but not fully comparable to typical software developers, a set of priorities in their software testing and quality assurance approaches.

A survey on the state of the practice in game software development in the Austrian games industry was reported in 2010. The survey showed that game developers apply ad-hoc and flexible development processes and there are limitations in support for systematic software engineering methods [13].

Another paper focused on the issue of requirements engineering in game development [14]. The authors argued that classical requirements engineering is not readily applicable to games. The evidence, synthesized in the paper, identified the need to extend traditional requirements engineering techniques to support the creative process in game development.

The video game industry is known for its intense work ethics with long hours of overtime, known as "crunch time". Crunch time involves periods of extreme workload, sometimes lasting several weeks, typically prior to final releases. Edholm *et al.* explored reasons and effects of crunch time by interviewing game developers from four different game studios [5]. The authors conclude that the creative passion of the developers opens up for crunch time, and the periods are initiated by unrealistic schedules and feature creep. Given the definition of "crunch time", a game jam would be an example of a short voluntary crunching.

## 3 RESEARCH METHOD

Our research method was questionnaire-based opinion survey which is an established method to gather empirical data in software engineering and other fields, e.g., [15, 16]. To design the survey, we used and benefited from survey guidelines as reported in the literature, e.g., [15, 17]. We also used our experience in designing and conducting several opinion surveys in the recent past, e.g., [18-20]. We discuss next: (1) survey design (how we designed the survey questions); (2) survey execution and sampling strategy; and (3) data analysis and synthesis approach.

### 3.1 SURVEY DESIGN

We first designed an initial version of the survey, considering the recommendation from the guidelines [13, 15] to keep the number of questions to a minimum – as survey participation understandably drops for longer surveys. Table 1 shows the structure of the questionnaire used for the survey, which has five parts and 16 questions in total.

**Table 1-An overview of the survey questionnaire**

| Part | Corresponding RQ / topics of questions | Num. of questions |
|---|---|---|
| 1-Background information | <ul><li>Country</li><li>Number of jams attended</li><li>Years of game development experience</li><li>Team size</li><li>Name of game engine(s) used (if any)</li><li>Role(s) in the team (choose any number from a provided list)</li><li>Work position (outside the game jam event)</li></ul> | 7 |
| 2-Expectations on the game (requirements) | <ul><li>RQ 1.1: How expectations were captured (freeform text)</li><li>RQ 1.2: How changed expectations were communicated within team (freeform text)</li></ul> | 2 |
| 3-Satisfaction assessment (QA and testing of the game) | <ul><li>RQ 2.1: QA during game development (freeform text)</li><li>RQ 2.2: QA at the end of the game development, i.e., how the final product was assessed w.r.t. expectations (freeform text)</li></ul> | 2 |
| 4- Software development practices | <ul><li>RQ 3:<ul><li>Software development practices applied (choose any number from a provided list with 22 items)</li><li>Level of familiarity with agile software development</li><li>Level of familiarity with lean software development</li></ul></li></ul> | 3 |
| 5-Concluding questions | <ul><li>Perception of success level in GGJ-2017</li><li>Any other comments</li></ul> | 2 |

Part 1 of the survey asked seven questions on background information, e.g., country of GGJ participation, number of game jam events attended, years of game development experience, and team size.

Three of the parts (2, 3, and 4) are for the study's three RQs (as raised in Section 1). In parts 2 and 3, we want to capture the participants' ways of working in an unfiltered and spontaneous way. Hence, the questions in these parts are open. To triangulate and relativize parts 2 and 3, we also have a part containing closed questions. Part 4 primarily consists of a list of development practices compiled from a mix of software development and game development sources. We condensed the list to 22 practices and an "Other (please specify)" option based on previous game development surveys [11, 21], established practices in agile and lean software development [22, 23], and an informal review of the game development blogosphere. The list of practices included items such as: continuous integration, minimum viable product, scope management, version control, stand-up meetings and paper prototyping.

Finally, part 5 asked two concluding questions – on an ordinal scale from "not familiar at all" to "extremely familiar" - specifically on familiarity of lean and agile practices. Lean and agile methods are contemporary and commonly used. Hence, we added two related questions to allow triangulation and improved validity of conclusions related to these practices. In addition to the questions on agile and lean, part 5 also included two general questions on the respondents' overall perception (on an ordinal scale) and a concluding freeform question for any final comments.

Once we had the initial version of the survey, we followed the advice of the survey guidelines [15, 17], and conducted a pilot phase with the survey's initial version among a set of participants from GGJ 2016. Based on 13 responses that we received in the pilot phase, we removed one question (name of the team) and added two new questions: country of the event, and name(s) of the game engine(s) used in game development. We also improved the wording (terminology) of some concepts to ensure clear and consistent understanding of the concepts asked in the survey questions, e.g., some respondents of the pilot phase mentioned that they were not familiar with the "software requirements" term (perhaps their backgrounds were from outside software engineering), thus we put both phrases "game expectations" and "software requirements" in the second part of the questionnaire (see Table 1).

Full versions of both versions of the survey questionnaire (initial and final), as presented to the participants, can be found in this online source [24].

## 3.2 SURVEY EXECUTION AND SAMPLING STRATEGY

Once the survey questionnaire was ready, we planned and executed the survey. To invite respondents for data collection, one of the authors attended a game jam in Malmö, Sweden on January 2017 ("Malmö Jams Too"). The author approached all teams present at the event in the afternoon of the second day, explained the purpose of the study, and shared a link to the online questionnaire.

Our sampling method was "convenience sampling". In convenience sampling, "Subjects are selected because of their convenient accessibility to the researcher. These subjects are chosen simply because they are the easiest to obtain for the study. This technique is easy, fast and usually the least expensive and troublesome. ... The criticism of this technique is that bias is introduced into the sample" [25]. As reported in a highly-cited survey of controlled experiments in software engineering [26], convenience sampling is the dominant survey approach in software engineering. Albeit its drawbacks and bias in the data, this does not mean that convenience sampling is generally inappropriate. For example, Ferber [27] refers to the exploratory, the illustrative, and the clinical situations in which convenience sampling may be appropriate. Convenience sampling is also common in other disciplines such as clinical medicine and social sciences, e.g., [28, 29].

To further publicize the survey, we asked the global GGJ organization team to post a tweet about our survey from the official GGJ Twitter account (@GlobalGameJam). We received 198 responses in total. To estimate the response rates, we had access to two numbers: (1) the total number of GGJ 2017 participants, and (2) the number of followers of @GlobalGameJam Twitter account, as shown in Table 2. The response rate based on these two sampling pools would be 0.54% and 1.15%, respectively. There were 36,401 registered participants for GGJ 2017, in 701 sites, and in 95 countries. A total of 7,263 games (accessible on https://globalgamejam.org/history) were developed in this event. The event took two full-days.

**Table 2- Response rates of the survey**

| Sampling pool (population) | Total population size | Response rate (upper bound) |
|---|---|---|
| Participants of GGJ 2017 | 36,401 registered participants | 0.54% |
| Followers of @GlobalGameJam Twitter account, as of Jan 21, 2017 | 17,300 followers[1] | 1.15% |

## 3.3 DATA ANALYSIS AND SYNTHESIS APPROACH

As Table 1 shows, parts 1, 4, and 5 primarily contained questions with predefined lists from which respondents could choose a value from (e.g., role in the team, and development practices applied), or were simple data such as integers or strings (e.g., country and number of game jams attended). However, responses to four of the survey questions were to be provided as freeform text which made them qualitative data. Two of those questions were related to RQ1 (Expectations on the game) and two were related to RQ2 (QA and satisfaction assessment).

To perform qualitative data analysis, we conducted "open" followed by "axial" coding [30] of the raw data as reported by the respondents. Open coding means inductively developing codes while reading the raw data. Subsequently, axial coding integrates and organizes the codes to construct linkages between data. The authors have had experience in this type of

---

qualitative data analysis in some other recent papers, e.g., [31, 32]. All authors were involved in the coding. The first and second authors were responsible for the open and axial coding related to RQs 2.1 and 2.2, whereas the third and fourth authors focused on RQs 1.1 and 1.2. We then compared the resulting codes and adapted them slightly for better alignment. The fifth author, a game development professional, had a supervisory role to validate the relevance of codes and any subsequent any conclusions.

Basically, we first collected all the factors related to all the four above questions in the dataset. Then we aimed at finding factors that would accurately represent all the extracted items but at the same time not be too detailed so that it would still provide a useful overview, i.e., we chose the most suitable level of "abstraction" as recommended by qualitative data analysis guidelines [30]. When necessary (e.g., there were too many items under one factor) and only if it made sense, we divided a factor further down, e.g., the "playtesting" group was broken down to internal playtesting and external playtesting.

For example, to answer RQ 2.1 (QA during game development), we did an initial screening of the provided freeform text responses and prepared an initial set of classifications, e.g., regular communications (among team members), and playtesting, which is the process by which a game designer tests a new game for defects and design flaws [33]. During the qualitative data analysis process, we found out that our pre-determined list of factors had to be expanded, thus, the rest of the factors emerged from the provided responses. The creation of the new factors (group) in the "coding" phase was an iterative and interactive process in which all the researchers participated.

Building on the same example, **Fel! Hittar inte referenskälla.** shows a snapshot of the qualitative coding to answer RQ 2.1 (QA during game development). This particular respondent had mentioned that "*to see if everything was following everyone's expectations, we would always meet up and play through the game*". The two phrases "*we would always meet up*" and "*play through the game*" led to this response being classified under the two items: regular communications, and internal playtesting, as shown in **Fel! Hittar inte referenskälla.**.

Figure 1-A snapshot showing qualitative coding of survey data to answer RQ 2.1. The green row shows the number of non-empty cells per column.

In a similar manner, qualitative coding of data for RQs 1.1, 1.2 and 2.2 were conducted and we will report those results in Sections 4.2 and 4.3.

On another issue, we have assigned anonymous IDs to the respondents and their data records (as seen in **Fel! Hittar inte referenskälla.**). In the rest of the paper, when we reference certain data from certain respondents, we will use the "Ri" format, e.g., we can see the data row of R3 in **Fel! Hittar inte referenskälla.**.

# 4 RESULTS AND DISCUSSION

We present the survey results and analyze them in the next several sections.

## 4.1 DEMOGRAPHICS

As discussed in Section 3.1, the demographics data include information about respondents' backgrounds and consisted of seven items: Country, Number of game jams attended, Years of game development experience, Team size, Contributions in the team, Position in actual work (outside the game jam), and Name of game engine(s) used (if any).

In terms of country of residence (work), developers from 51 countries filled out our survey. The five countries with the highest number of respondents were: France (31 respondents), United States (17), Brazil (11), Germany (9), and Poland (7).

The six charts in Figure 2 represent the survey data corresponding to the above demographics. In terms of number of game jams attended, the histogram in Figure 2 shows that the trend is left skewed, thus most respondents had attended a few game jams as of the survey. The average of the number of game jams attended was 3.2.

In terms of years of game development experience, many respondents had between 1-4 years of experience and then many had less than 1 year experience. Only two of the 198 respondents had more than 15 years of experience in game development.

In terms of team size, five was the most frequent size. Seven respondents mentioned that their teams had more than 10 members. The average of the team sizes was five.

In terms of contributions in the team, most of the respondents were involved in the programming task. Many other respondents also contributed art, design, and audio artifacts to the game projects.

In terms of job position outside the game jam, there was a wide variety in responses. Many respondents were students of game development, or had game/software developer titles. Also, a few teachers (instructors) were present.

We were also curious about the game engine(s) used. Unity (also named Unity3D), www.unity3d.com, is by far the most widely used engine in the GGJ (used by 78% if the respondents). This came as no surprise since Unity Technologies, the corporation behind the game engine, has been sponsoring the event for several years now.

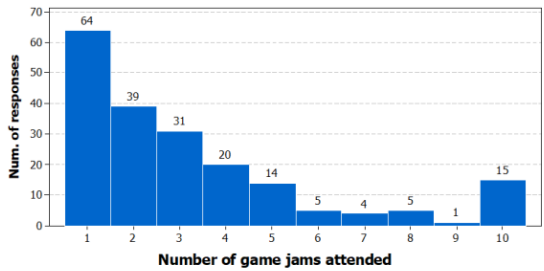

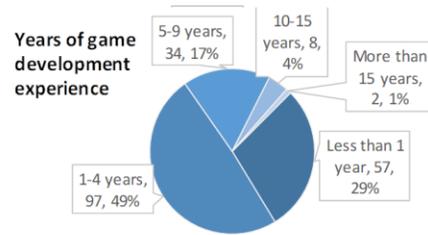

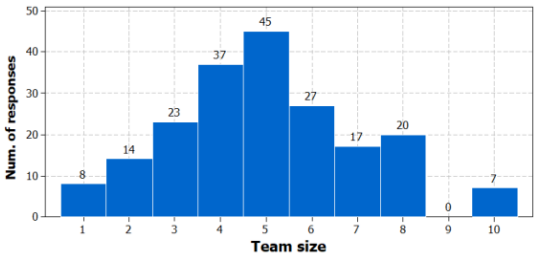

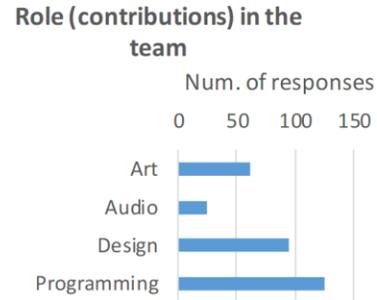

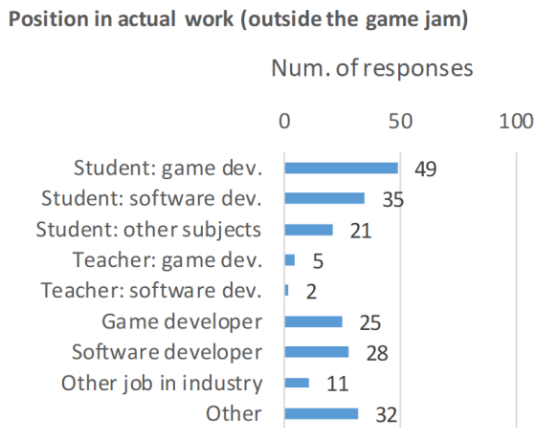

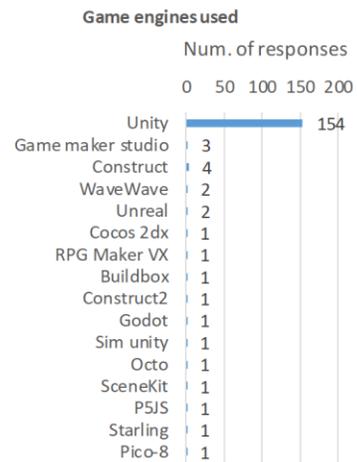

**Figure 2- Demographics data of the survey respondents**

Also, as a concluding question, we asked about the respondents' perception of success level in GGJ 2017 in a 5-point ordinal scale (1-5), in which 1 would denote "Not successful at all" and 5 would mean "Extremely successful". Figure 3 shows the histogram of the perceptions, which is right-skewed, denoting that the majority of respondents reported that their team was successful at the game jam.

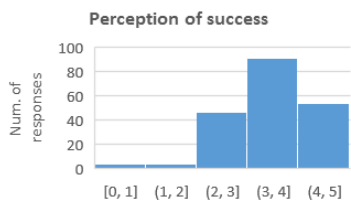

**Figure 3- Perception of team success in GGJ 2017**

## 4.2 RQ 1-GAME "EXPECTATIONS"

RQ 1 had two sub-RQs which we discuss next:

- RQ 1.1: How is initial gathering of game expectations conducted?
- RQ 1.2: How is evolution and change of game expectations handled?

### 4.2.1 Initial gathering of game expectations

To get a sense of how the initial expectations were gathered, we asked the following open-ended question: How did your team initially capture expectations (requirements) of the game, and how did you ensure that all team members shared the same understanding?

A total of 187 freeform answers were collected in relation to RQ 1.1. As discussed in Section 3.3, we conduced qualitative data analysis (coding) to categorize the freeform data. The results, shown in Figure 4, indicate that holding iterative sessions of brainstorming (76%) was the main method used for conceptualizing the initial set of requirements, or expectations. This was the dominating theme in most of the answers. For example, *"Spent some time brainstorming as a group & came up with several ideas, then discussed pros & cons of each & settled on one"* and *"We brainstormed and decided on the best idea"*.

Several groups have reported using a whiteboard (6%) for sketching out their original set of ideas to make sure that all team members were on the same page. For example, *"We put a ton of vague concepts on the whiteboard and started slapping them together to make game ideas"*. Several teams have reported that their main ideas originated from preconceived ideas (8%), such as attending prior game jams or based on existing video games that some of the team members used to play. For example, one respondent (R30) stated that, *"Most of us attended to previous game jams, so we knew the kind of expectations we could put on the game"*.

Majority voting (3%) and prototyping (3%) were used by a small number of teams to determine which ideas to proceed with and which ideas to drop. For example, one respondent (R23) indicated that, *"We brainstormed then explained then voted on the games we liked the most"*, and another respondent (R6) stated that, *"Group Brainstormed, roughly prototyped ideas, scoped small"*. Prototyping was mainly used to ensure that the ideas were technically feasible as requirements. Finally, a few teams (4%) had only one member with either a preconceived idea or the single person made all the requirements calls during the event. For example, respondent R86 said that: *"Since it was only me in the team, I was able to do everything that I wanted according to my expectations"*..

Our analysis also shows that a major factor that influenced decision–making during this phase was the fact that most participants were familiar with the time and technology constraints of the game jam. Therefore, they kept their expectations as realistic as possible. This was detected in answers such as, *"We agreed on a much smaller project than last year"* (R33) and *"We were pretty sure what we can do in just two days and knew what we wanted to reach"* (R40).

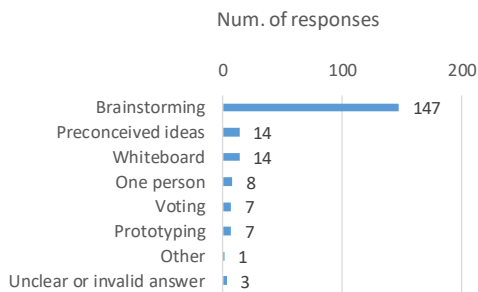

**Figure 4-Gathering the initial set of expectations (requirements)**

#### 4.2.2 Evolution and change of game expectations

There were 178 freeform answers to the question on how the team communicated around changes in the expectations, i.e., 90% of the respondents answered this question. Figure 5 shows the results of qualitative analysis (coding) of those data.

Interestingly, 16 of the 178 answers (9%) answered that their expectations did not change. For example, R70 states that *"We stayed close to the main idea"* and R92 *"We didn't prepare the expectation in detail so there were no real changes"*. We interpret those comments as that the overall expectations did not necessarily change but they were refined and detailed. Other respondents, however, were more direct and answered *"The expectations did not change"* (R81) and *"No, the expectations did not change"* (R93).

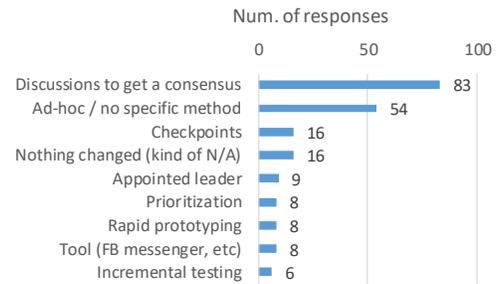

**Figure 5-How changes to the expectations were handled**

The most common way – 47% of the answers - the teams handled changes were to discuss among the team members and reach a consensus. For example, R16 answered that *"We were working in close quarters. There was a constant conversation on the direction the game was going"*. This is quite expected since the small teams are in fact co-located and know each other from before. Similarly, 16 respondents (9%) reported using checkpoints – to a varying degree of formalism - where the team would gather and compare progress to the plan. E.g. R84 stated that *"We rest for lunch and dinner every day and reevaluate our targets and focus."* and R144 *"Every three or four hours, we did a short stand-up meeting to see where we are and what to do next"*. There is no significant correlation with experience or any software practices, indicating that direct communication within a team of equals is intrinsic to this kind of collaborative work.

Nine respondents (5%) reported having an appointed leader who decided on any changes. R89 reported that *"We had small discussions and I decided. We did not want to lose time."* and R173 *"Programmers would bring up challenges and the lead designer would make changes accordingly and communicate it with relevant parties"*. Eight respondents (4%) responded that there was prioritization being performed through a voting. R34 mentioned *"By calling everyone and explaining what happened, and voting on whether or not to change plans."* and R172 *"Mini brainstorm sessions and a vote on what could and couldn't be executed in the time frame"*.. Hence, it seems quite few teams (5%) had a strong leader making most of the decisions and the rest of the teams either discussed or voted to reach an agreement.

Tool usage is scarcely reported; only eight respondents (4%) mentioned one or more tools. Examples which are mentioned are Trello, Slack and Facebook messenger. Given that the team size varies between 4-6 individuals (from the lower quartile to the upper quartile) in combination with the time pressure, it is hardly surprising that there is a lack of tool usage. We hypothesize that this can be significant for larger endeavors as well in that it is important to have simple and readily available tools, especially close to a deadline rather than comprehensive and complicated ones.

### 4.3 RQ 2-QA AND TESTING "EXPECTATIONS"

RQ 2 had two sub-RQs which we discuss next:

- RQ 2.1: What type of QA is done during game development?
- RQ 2.2: What type of QA is done on the final product (game)?

### 4.3.1 QA during game development

There were 164 freeform responses to this question, i.e., 83% of the respondents answered this question. Figure 6 shows an overview of the results.

Forty-one respondents (26%) reported having regular communications to ensure whether the development progressed in line with the shared expectations on their games. The next popular ways for QA were, in order: internal playtesting, dynamic and proactive planning, and clear definitions of individual tasks. It is interesting that regular communication is ranked higher than other "technical" QA activities, such as (internal or external) playtesting.

Regular communications among team members was an important QA activity. For example, R35 mentioned that "*We constantly spoke about the realistic progress of things, speaking about where we wanted to take it from there. As design I had a very clear road map which I identified and explained to my team, however personal desires and opinions sometimes were better or more validated or just something easy to throw in for shits and giggles*". R66 said that: "*We talked*". For QA, R97 said that: "*Open communication, every couple of hours. Making sure everyone was engaged was key*".

Internal playtesting is also popular, which is the process by which a game designer tests a new game for defects and design flaws [33]. Playtesting is an active topic in practice, e.g., [34, 35], and also an research area in the game community, e.g., [36-39]. Playtesting could be done by members of the development team or external testers, which we refer to as external playtesting. Internal playtesting (mentioned in 33 responses) was more popular than external playtesting (only 2 responses), which could possibly be due to the time pressure (shortage) in the context of GGJ, or that a given team did not know other teams well and thus did not ask them to playtest their game during development. Here are some example quotes on using playtesting as a QA approach: R3 said that "*to see if everything was following everyone's expectations we would always meet up and play through the game*". R16 said: "*We play tested the game together every four hours*". R34 said: "*Several playtests, including using people from other teams as beta testers*" referring usage of both internal and external playtesting.

Dynamic and proactive planning was also a popular QA approach. For example, R25 said: "*As time passed by, we'd analyze what tasks were yet to be completed and would remove anything with a big chance of extrapolating the time limit*". R163 mentioned that their team set "*short milestones and reviewed them constantly*".

Eighteen respondents set clear definitions of individual tasks to ensure QA. For example, R10 said: "*We were so lucky as to have very set tasks for each member of the team, one person were assigned to only keep track of progress and time lines*". Similarly, R38 mentioned: "*Proper task management: letting team members know what everyone was working on and what else still needed to be done*".

Focusing on developing main features first (prioritization of features) was another QA approach, mentioned by 11 respondents. Dynamic feature planning and prioritization is a popular approach in software engineering (especially in Agile methods) [40, 41]. Here are some examples from the dataset. The team in which R73 was a member of "*implemented main mechanics early on and tested them*". R123 said that their team "*agreed on what the main functionality expected of the game was. All the ideas that didn't fall in there were kept in a separate list to implement if there was enough time*". R139 said that: "*Our team could see the progress in real-time as we built the foundations, and the* rest of the game upon them part by part until we complete everything planned".

Another QA mechanism was to decrease the scope / expectations. For example, R43 said: "*we saw that we needed more art and programmers and for that we had to cut down the expectations of the project*". R75 said: "*when progress is [was] behind, we dropped some elements [features]*". R184 said: "*In a nutshell, we had to remove certain aspects from the game to meet time requirements*".

11 respondents mentioned that their teams had dedicated team members for QA. For example, R20 said: "*Only one member did [the QA], and berated others in line*". R40 said: "*The producer had that task, he knew how we were doing*". R58 said: "*I worked as producer for my team, so I gave direction and kept everyone on the same track with frequent check-ins*".

Several "Other" QA activities were also mentioned, e.g., involving designers in QA (R4), prototyping (R111 and R116), and minimum viable product (R75). One respondents (R55) explicitly mentioned that: "*We had not much time to do that [QA]. We rushed to get something done*".

Unfortunately, for this question in particular, there were many unclear, vague or invalid answers (48 of the 164 responses=29%), which we found not related to issue under study (QA during game development). Some of those responses were: "*Just ok*" (R9), "*It was hard to evaluate if we were in time with where we needed to be because it was my first jam*" (R13), "*I was effectively the only member of my team*" (R28), and "*we made what we mentioned to made*" (R29).

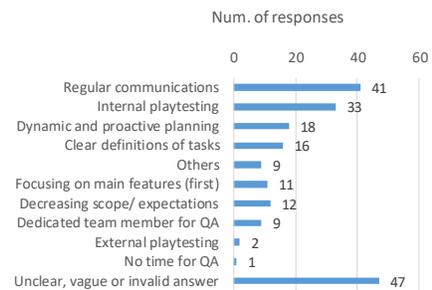

**Figure 6- QA during game development**

### 4.3.2 QA of final product

The next question in the part on "satisfaction assessment" was answered by 171 respondents (86% of the respondents). Since the respondents could leave any of the four consecutive freeform questions blank, the relatively high fraction suggests that the questionnaire did not overwhelm the respondents. Figure 7 shows the results from our qualitative coding.

Internal playtesting was by far the most common approach to do concluding QA before the GGJ deadline, reported by 28% (48 out of 171) of the respondents. As for RQ2.1 on QA during game development, we believe that this approach to QA is popular when there is considerable time pressure – as during the final hours of a game jam. Example responses include "*Played it a lot. If we enjoyed it, we assumed others would too*" (R118), "*We tried to beat our own scores on the game we just made*" (R162), and "*very quick playtests*" (R198).

External playtesting is considerably more popular as a QA approach for the final product (18 out of 171, 10%) compared to during ongoing development. We put forward two possibly complementary explanations. First, groups might prioritize getting external input on the game in the final stage of a game jam as R181 explained: "*We had other jam participants to play our game to find if the game was good enough. Also we had some game development professionals come to the jam site so seeing that they enjoyed the game was great*". Second, groups might be ready well before the deadline and decide to use the remaining time to let other available GGJ participants provide concluding feedback. R50 represents a group that finished early: "*we finished the game 10 hours before the deadline, we dedicated the whole 10 hours exclusively to playtest and fix smaller issues*".

We notice that a bare minimum of QA, or even no concluding QA at all, are common responses. Twenty-one respondents (12%) report that they did no concluding testing prior to the deadline. Several respondents refer to the time pressure, e.g., "*We did not assess much, because we were behind schedule and crunched in order to deliver a playable game.*" (R4) and "*By grinding to get the gameplay loop closed. We didn't have enough time to balance and add polish and any squishiness*" (R35). Seventeen respondents explained that their team had minimal expectations on the resulting game, and that anything playable would constitute a success. Eleven respondents report that they only did testing of fundamental game mechanics, e.g., "*We simply ensured that we had a playable build*" (R60) and "*We just focused on minimal playable delivery first to be sure having something to show*" (R106).

Regular communication, the most common approach to QA during GGJ development, was also mentioned by 11 respondents in relation to QA of the final product. It is evident that communication throughout the development can support final stage QA, as explained by R145: "*The same way we did throughout the whole game jam, through frequent communication. Although it was harder in the end because of last minute rush situations, I do think it would have been a lot worse had we not taken the time*". However, we believe that more than 11 respondents were part of groups that practiced regular communication – if it was mentioned in the previous question, it is likely to apply until the end of the game jam.

Several respondents had an approach to concluding QA that relied on careful previous planning. Fourteen respondents (8%) stressed that the quality was good enough when all specified work items were completed. Respondents mentioned Kanban boards and master development tables: "*By comparing what was done with what remained to be done, based on our Kanban board.*" (R144) and "*We just checked if the tasks were completed in the development master table.*" (R19). Eight respondents (5%) report that they did not complete all tasks they had planned for.

A subset of the respondents reported concluding QA geared toward the elusive concept of "feel" and emotions. Twelve respondents (7%) focused on assessment of game feel such as "*in the last hours we used more gut feeling that structured testing*" (R84, sometimes in combination with internal playtesting: "*[we did QA through] playtesting and then discussing what feels wrong and what feels right*" (R99). Nine respondents (5%) mentioned that they used comments and reactions from the GGJ audience and by walkers as an approach to concluding QA, e.g., "*by the other jammers' reactions.*" (R53) and "*People on our site gave us good feedback and many loved our concept.*" (R33).

As for the previous questionnaire question, there were several responses that we could not code due to unclear or vague content. However, the fraction was lower (24 out of 171 responses, 14%), suggesting that the questions was easier for the respondents to understand. Examples include: "*It is as expected but it lacks in optimization*" (R37), "*We still think that the quality of the game is so good especially if we consider time we have spent at it*" (R82), and "*Pretty well but lacked gameplay mechanics*" (R83).

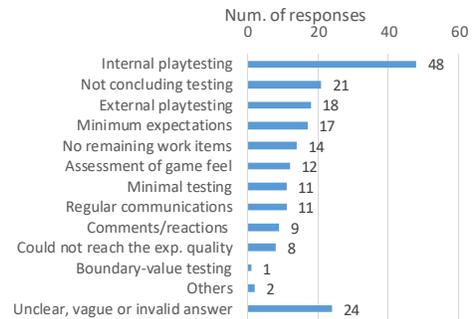

**Figure 7- QA of the final product (game)**

## 4.4 RQ 3-APPLICATION OF SOFTWARE DEVELOPMENT PRACTICES

Almost all respondents (195 out of 198, 98%) provided responses to this question by choosing one or more software development practices from a provided list (Section 3.1).

Figure 8 shows the results. The top five used software development practices were: continuous integration, minimum viable product, scope management, version control, and stand-up meetings. Continuous integration, even without dedicated tool support on a server, is a sensible practice when there is no time to resolve problems from late "big bang" integration. Furthermore, both minimum viable product and scope management, i.e., two project management practices, are expected results as they mitigate the risk of not having a final game to deliver in the end of the game jam. Version control and stand-up meetings are interestingly widespread, a technically-oriented practice and a human-oriented practice, respectively. Both practices could introduce some overhead, i.e., setting up a version control server and stopping all development activities for a short time to discuss current status, but apparently participants believe that their values exceed the costs.

We found it somewhat surprising that automated testing and static code analysis are rarely used practices. The reason is most probably due to the time pressure, as one would not value automated testing or static code analysis for a development project with duration of two days only. Another observation is that exploratory testing, i.e., testing software without pre-designed test cases by intermixing design, execution, and analysis of tests guided by increased understanding of the subject under test [42]. However, we suspect a terminology mismatch, as we believe the "playtesting" frequently reported in relation to RQ2 is typically exploratory in nature – rather than guided by pre-designed (thus repeatable) test cases.

We also calculated the number of software development practices applied by each respondent. Results are also shown in Figure 8. We can see that most respondents reported applying many development practices, suggesting that the respondents were knowledgeable of the domain.

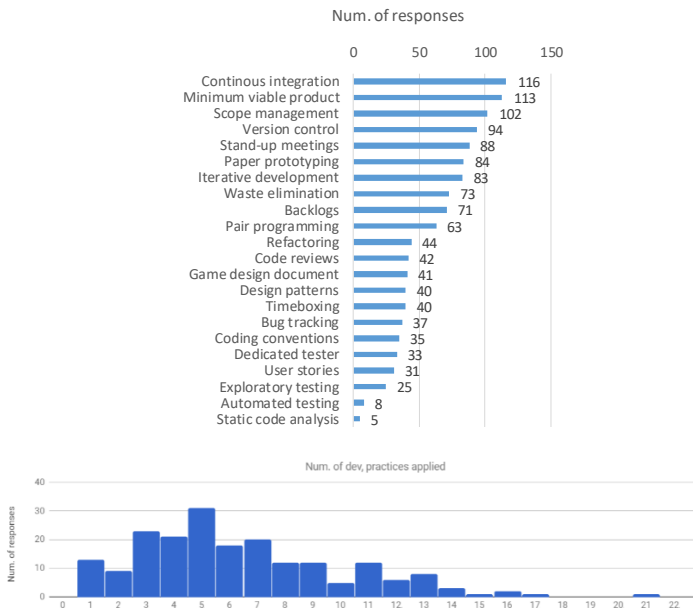

**Figure 8- Application of software development practices**

## 4.5 RQ 4-CORRELATIONS OF DATA

We hypothesized that various factors may have cross-correlations. We explored the possible correlations among the following parameters: familiarity with lean, familiarity with agile development, team size, number of development practices applied, and perception of team success in the GGJ. Table 3-Correlations of data

| | Familiarity with agile dev. | Familiarity with lean dev. | Perception of success | Team size |
|---|---|---|---|---|
| Familiarity with lean dev. | 0.57 (Pearson coefficient), 0.00 (p-value)  0.57 (Spearman's rho), 0.00 (p-value) | | | |
| Perception of success | 0.27, 0.00  0.31, 0.00 | 0.19, 0.01  0.21, 0.00 | | |
| Team size | -0.09, 0.19  -0.09, 0.17 | -0.02, 0.73  -0.02, 0.67 | -0.06, 0.37  -0.04, 0.57 | |
| Num. of dev. practices applied | 0.27, 0.00  0.31, 0.00 | 0.18, 0.01  0.20, 0.00 | 0.25, 0.00  0.28, 0.00 | 0.16, 0.02  0.14, 0.03 |

shows the Pearson and Spearman correlation coefficients of those factors along with p-values.

We are showing the weak to high correlations in underlined fonts. As we can see in the data, familiarity with lean has modest correlation (coefficient=0.57) with familiarity with agile development, as expected, since if one is familiar with one of these concepts, s/he would be familiar with the other. Interestingly, familiarity with agile development and the number of development practices applied both have weak correlation with perception of team success in the GGJ. Our results suggest that GGJ participants that are familiar with agile development, and apply several development practices, tend to perceive the game jam as successful. However, whether this finding relates to a delivery of a better game in the end or a more harmonious team experience during the game jam requires further research. Furthermore, we emphasize that the self-assessed perception of success is a highly individual measure, relative to the individual's personal goals at GGJ – a first-timer might be satisfied regardless of the result, whereas a professional developer might enter GGJ with very high expectations.

## 5 IMPLICATIONS OF THE RESULTS

In this section, we discuss the practical implications of the results from our survey. First, we discuss aspects that might be valuable to future game jam participants, especially first timers. Second, we discuss the results in the light of time pressure in game and software development in general.

**Table 3-Correlations of data**

| | Familiarity with agile dev. | Familiarity with lean dev. | Perception of success | Team size |
|---|---|---|---|---|
| Familiarity with lean dev. | 0.57 (Pearson coefficient), 0.00 (p-value)  0.57 (Spearman's rho), 0.00 (p-value) | | | |
| Perception of success | 0.27, 0.00  0.31, 0.00 | 0.19, 0.01  0.21, 0.00 | | |
| Team size | -0.09, 0.19  -0.09, 0.17 | -0.02, 0.73  -0.02, 0.67 | -0.06, 0.37  -0.04, 0.57 | |
| Num. of dev. practices applied | 0.27, 0.00  0.31, 0.00 | 0.18, 0.01  0.20, 0.00 | 0.25, 0.00  0.28, 0.00 | 0.16, 0.02  0.14, 0.03 |

## 5.1 CONTEMPORARY GAME JAM PRACTICES AND IMPROVEMENT SUGGESTIONS

The primary finding from the survey is that the time pressure characterizing game jam participation stimulates *ad hoc* approaches to video game development. This is an expected finding, as a game jam is a one-shot project with an extremely challenging deadline. Brainstorming dominates as the approach to elicit game expectations, rarely supported by anything more advanced than pen and paper or whiteboards. As expectations evolve during the game jam, the participants tend to meet and discuss to reach agreement. Some teams also reconvene at regular checkpoints, in line with recommendations on how to coordinate work in co-located agile development [43]. Face-to-face communication appears to be sufficient for most teams, but some respondents report using Trello, Slack, and Facebook Messenger to communicate changed expectations.

Regular communication is not only the main approach to align expectations within teams at a game jam, it is also the primary means to monitor that the game under development will meet the expectations. Another common approach to QA is that team members playtest their game themselves, an approach whose value can be further increased by efficient communication within the team. Research shows that human factors such as communication and cooperation are critical to alignment of expectations and testing in large software development projects [44], but the *ad hoc* nature of game jam development might further amplify the importance of communication – as there simply are few alternatives, i.e., there are typically no artifacts, tools or documents for team members to refer to.

Several game jam participants refer to successful planning as an approach to QA, especially being dynamic and proactive in the planning process. Some teams plan and specify development tasks with clear "Definitions of Done" (Scrum terminology) to embed QA in the tasks. Other teams, with a focus on planning, regularly manage the scope of game features to make sure that what remains in the final game meets the expectations. While decreasing scope is a natural activity in agile development, game studios are reportedly bad at this despite claiming to have adopted agile practices, thus contributing to periods of crunch time [5]. During game jams, teams are already in a crunch mode, and might proactively descope features to support the quality of the final game – at the same time mitigating even more intensive crunching.

When it comes to QA of the final version of the game, playtesting within the team is the most common approach. Some teams also use feedback from other game jam attendees, either letting them playtest the game or by collecting impressions from bywalkers. To what extent collecting external input on the final game is a strategic decision, or if it rather happens opportunistically, cannot be deducted from our data. What we can say, however, is that there exist game jam teams that prioritize playtesting over implementation of new features as early as 10 hours before the deadline – indicating high quality standards. On the other hand, many teams also report that they had no time for any concluding QA prior to the deadline.

A game jam is a period of crunch time, i.e., the context requires the participants to focus the available time on activities that directly provide value for the final game and skip everything else (in line with lean software development). What practices remain when in such a rush? Interestingly, the GGJ17 participants report using many of the game and software development practices listed in the questionnaire. At the game jam, the five most common practices are: continuous integration, minimum viable product, scope management, version control, and stand-up meetings – indicating their value under extreme time pressure.

On the other side of the spectrum, static code analysis and automated testing is hardly used at all. As there is a considerable initial investment to get started with automated testing [45], apparently game jam participants do not perceive the cost-benefit analysis as positive. Open source static code analysis tools, on the other hand, could be installed before a game jam – thus we consider this an idea for game jam participants to explore in the future, to provide support for identification of common code smells and potential bugs [46], e.g., poor memory management.

Several development practices that have been empirically shown to support software development are used by less than a third of the game jam participants. Refactoring has been shown to support both software quality and developer productivity for small teams working in highly volatile domains [47], i.e., a context that shares similarities with a game jam. Code review is an established practice to promote quality and stimulate knowledge transfer within a team [48]. Since modern code reviews practices are lightweight (and supported by mature open source tools), they could possibly be encouraged during a game jams to support communication. Finally, adhering to coding conventions, e.g., identifier naming and formatting, supports source code readability that can help to surface bugs [49]. Agreeing on a simple coding convention would not require additional effort by a game jam team, instead it could help the developers focus their creativity elsewhere.

## 5.2 GENERALIZING TO TIME PRESSURE BEYOND GAME JAMS

We argue that the extreme time pressure a game jam team experiences is an interesting phenomenon to view also from a broader software development lens. From our perspective, the development context closest to game jams that has been targeted by academic research is software startups. A software startup resembles a game jam team as it also faces intense time pressure and tends to operate with a small number of developers in a chaotic, rapidly evolving context with substantial uncertainty. A systematic mapping study by Paternoster *et al.* shows that time pressure is a recurrent theme in the literature as "the environment often forces startups to release fast and to work under constant pressure" [50].

Paternoster *et al.* also present a collection of practices that have been reported to be useful in startups, and several are in line with our findings/recommendations for game jams (our findings in parentheses). Example include: (1) lightweight development methods that offer flexibility (*ad hoc* development), (2) evolutionary prototyping (minimum viable product, iterative development), (3) simple requirements engineering to reduce waste (game design document, backlog, waste reduction), (4) relying on well-known frameworks (Unity), and (5) use of open source software tools (code reviews, static code analysis). Finally, Paternoster *et al.* report that (6) software testing and QA is often compromised in startups and recommend using early adopters as a QA team (external playtesting) or outsourcing QA to experts to let the developers focus on programming (dedicated tester). Due to the many contextual similarities, we hypothesize that the ranking of development practices used at game jams (cf. Fig. 8) would be valid also for software startups – even more so for video game development startups.

Two other development contexts that display similarities to the time pressure of game jams are web development and mobile app development. The number of web development projects exploded during the dot.com era, and short cycle time software development or "Internet Speed Software" became the norm for web companies acting on the competitive Internet market [51]. For many companies, the time pressure turned quality into a negotiable concept. Rapid prototyping and "hacking" became tools to survive the competition, inevitably introducing technical debt. The present-day counterpart of the early web development is mobile app development. Developing apps involves rapid development on top of established platforms with pressing time-to-market expectations under fierce competition. Joorabchi *et al.* report that software testing for mobile apps is perceived as one of the biggest challenges by app developers, and that practitioners call for better tool support [52]. A key difference between both the time pressure of web and app development compared to game jams is that the time pressure continues after release, i.e., the users expect a steady stream of patches and do not accept waiting for bug fixes. At a game jam, however, the time pressure ends abruptly when the event is over – the incentives for game jam teams to prioritize maintainability of the software is thus very low.

Studying time pressure from a theoretical management perspective, Austin developed a model showing that a tight deadline actually can improve software quality [9]. He argues that "shortcuts" (similar to technical debt) are introduced when a developer makes implementation decisions to stay on schedule rather than developing software in line with the best interests of a project. If deadlines are leisurely set, a project will experience less shortcut-taking by its developers. On the other hand, if projects set deadlines that are harder to meet, the frequency of shortcuts increases. However, if projects regularly set aggressive deadlines that developers know are unachievable, then the stigma associated with missing internal deadlines is removed and developers again focus on quality of their individual project contributions.

Missing internal deadlines does not appear to be acceptable in video game development though. The widespread periods of crunching suggest that the stigma remains powerful enough to push developers to their limits. Peticca-Harris *et al.* refer to this as a "neo-normative control mechanism" [4], used to harness the passion of game developers and their portfolio-based reputations in the project-oriented (gig) economy. The popularity of GGJ shows that many people enjoy the atmosphere and sense of urgency – and thus voluntarily enter a 48-hour crunch. While we certainly support the organization of game jams, we also raise the issue that extreme work is a systemic problem in video game development. When crunching becomes the rule rather than the exception, sustainable working conditions are not provided, which might contribute to stress-related health problems. Moreover, crunching might oppose calls for diversity in the game development workforce, as fresh graduates might accept it to enter the business whereas workers with families must prioritize work-life balance differently. Game jam organizers are in a good position to teach representatives of the next generation of game developers that crunching is a rare event, thus mitigating the normalization of long hours of uncompensated overtime.

## 5.3 LIMITATIONS AND THREATS TO VALIDITY

The size of the population under study is known, i.e., the number of GGJ 2017 participants. The 198 responses correspond to a response rate of 0.5%. The primary means to invite respondents to the survey was through a tweet from the official GGJ Twitter account. Using Twitter might have introduced a bias towards GGJ participants active on this social media platform. However, as Twitter is both an official GGJ channel and popular among contemporary game designers, we believe that the sample is representative of the population.

A limitation from using Google Forms is that this questionnaire engine does not record partially completed surveys. Hence, we do not know the dropout rate. However, the responses from the pilot run suggested that the questionnaire was sufficiently relevant and engaging – and sufficiently concise.

*Content validity* concerns how much a measure represents every element of a construct. We had to ensure sufficient coverage of elements related to potential effects of time pressure in our questionnaire, but there is a trade-off between the length of the questionnaire and the coverage. To increase the number of complete answers, we opted for a relatively short questionnaire, with four open-ended questions targeting requirements engineering and software testing, complemented by a closed question for development practices. We deliberately excluded topics such as: software architecture, group dynamics, level design, and game art – all these would be worthwhile to explore in future work. Still, we believe that the threats to content validity are small as the answers indicate that we covered the relevant elements.

*Construct validity* refers to how an operational definition of a variable reflects the theoretical meaning of a concept. The major threat to our survey is whether our inquiry on game development under time pressure was properly interpreted by the respondents. Our preunderstanding is predominantly from software engineering research, which might have introduced a terminology mismatch, despite our efforts to avoid terms such as requirements engineering and QA. While parts 3 and 4 of the questionnaire did collect many invalid answers, we believe that this is expected as online open-ended questions are more difficult to get useful answers from. The successful instrument evaluation we conducted, i.e., the pilot run, supports our interpretation, along with one co-author being a game development professional.

*Internal validity* covers the degree to which causal relationships can be made, and to which extent confounding factors are at play. While our study is primarily exploratory, the opposing goals of GGJ teams might have influenced the collected answers. While many GGJ teams want to deliver the best possible game, others have different goals, e.g., experimenting with a new approach for learning purposes, or simply to enjoy themselves. As we did not control for different goals, our exploration of development practices that could support creation of the best possible game under extreme time pressure might be diluted by more easygoing respondents. Before presenting any game jam recommendations in future work, we will investigate this further.

Finally, *external validity* concerns the generalizability of our results. The response rate is low and participants from France are slightly overrepresented. Nonetheless, based on the distribution of backgrounds, roles, geographical location, and work situation, we believe that we have reached a representative sample of the population. Furthermore, we generalize the results using analytical generalization [53] rather than statistical generalization. Another threat to the external validity of our conclusions is that our questionnaire was geared at technical aspects of programming and software development. This might have created a bias toward GGJ participants focusing on game programming, thus challenging generalization to all roles of GGJ participants. However, as the respondents are distributed over different roles (cf. Fig. 2), this threat appears to be minor. Finally, regarding generalization of the time pressure at GGJ to other contexts, we emphasize that there are different types of crunch. Most importantly, crunching at a game jam involves the entire development process, but crunching in the video game industry is dominated by implementation tasks. Further research is needed to determine how this difference influences the experience of crunch time.

## 6 CONCLUSIONS AND FUTURE WORKS

A video game development project is a challenging endeavor, no matter if run by a small indie studio or a large organization with hundreds of developers. In the game industry, several established game and software engineering practices are applied to support development and delivery of high-quality game titles. However, game studios often work with ambitious time-to-market targets. Consequently, game programmers, designers, and artists often work under considerable time pressure. A frequently reported scheme to cope with pressing deadlines in the game industry is "crunch time", i.e., periods characterized by long hours of overtime. During a game jam, participants voluntarily enter a period of crunching to develop a game within a very short time span.

We conducted a survey of Global Gam Jam (GGJ) 2017 participants to explore how game jam teams tackle the time pressure. Our goal was to identify which practices remain when in a constant crunch time during an entire project. Based on 198 responses, we conclude that the time pressure stimulates *ad hoc* development approaches and lightweight methods to communicate and collaborate within teams. More specifically, the survey shows that: (1) iterative brainstorming is the most popular method for conceptualizing initial expectations; (2) continuous integration, minimum viable product, scope management, version control, and stand-up meetings are the most frequently applied development practices; (3) regular communication, internal playtesting, and dynamic and proactive planning are the most common quality assurance activities, and (4) participants' perception of team success at GGJ is weakly correlated with both familiarity of agile development and the number of applied development practices.

We encourage game jam teams to embrace lightweight methods and direct communication, but additional practices from the literature might increase the success rate of teams. Example practices with limited overhead include code reviews, refactoring, and use of static code analysis tools. Moreover, by comparing our findings to recommendations for software startups, we posit that some effects of time pressure on development during a game jam might be generalized to the startup scene – especially video game startups. Finally, we highlight that the crunching culture in the game industry is a concern and that game jam organizers are in a good position to problematize the phenomenon.

Our future work directions include the following: (1) exploration of additional topics in the game jam context, such as software architecture, level design, game art, and group dynamics including team morale, stress, and interpersonal relationships; (2) a systematic comparison of the contextual characteristics of game jam development and software startups, especially video game startups; (3) development and evaluation of guidelines for successful game jam development, and (4) collecting recommendations for how game jam organizers could incorporate a pedagogical mission to raise awareness about the negative sides of crunching – to support sustainable development in the video game industry.

## ACKNOWLEDGMENTS

The authors would like to thank the participants of Global Game Jam (GGJ) who participated in our survey.